\documentclass[aps,prb,twocolumn,superscriptaddress,showpacs,showkeys,floatfix]{revtex4}

\usepackage{graphicx}
\usepackage{dcolumn}
\usepackage[pdftex]{color}
\usepackage{maybemath}
\definecolor{darkgreen}{rgb}{0.133,0.545,0.133}
\definecolor{orange}{rgb}{1.0,0.76,0.02}

\begin{document}

\title{
{\rm\small\hfill (submitted to Phys. Rev. B)}\\
Diffusion of hydrogen within idealised grains of bcc-Fe: \\ A kinetic Monte Carlo study}

\author{Yaojun A. Du}
\affiliation{Fakult{\"a}t f{\"u}r Physik and Center for Nanointegration (CENIDE), Universit{\"a}t Duisburg-Essen, Lotharstra{\ss}e 1, 47048 Duisburg, Germany}
   
\author{Jutta Rogal}
\email[]{jutta.rogal@rub.de}
\affiliation{Interdisciplinary Centre for Advanced Materials Simulation, Ruhr-Universit{\"a}t Bochum, 44780 Bochum, Germany}

\author{Ralf Drautz}
\affiliation{Interdisciplinary Centre for Advanced Materials Simulation, Ruhr-Universit{\"a}t Bochum, 44780 Bochum, Germany}

\date{\today}

\begin{abstract}
Structural defects in materials such as vacancies, grain boundaries, and dislocations may trap hydrogen and a local accumulation of hydrogen at these defects can lead to the degradation of the materials properties.  An important aspect in obtaining insight into hydrogen induced embrittlement on the atomistic level is to understand the diffusion of hydrogen in these materials. In our study we employ kinetic Monte Carlo (kMC) simulations to investigate hydrogen diffusion in bcc iron within different microstructures.  All input data to the kMC model, such as available sites, solution energies, and diffusion barriers are obtained from first-principles calculations.   
We find that hydrogen mainly diffuses within the interface region with an overall diffusivity that is lower than in pure bcc-Fe bulk.  
The concentration dependence of the diffusion coefficient is strongly non-linear and the diffusion coefficient may even decrease with increasing hydrogen concentration.
To describe the macroscopic diffusion coefficient we derive an analytic expression as a function of hydrogen concentration and temperature which is in excellent agreement with our numerical results for idealised microstructures. 
\end{abstract}

\pacs{75.50.Bb,81.40.Np,67.63Gh,67.80.fh,88.30.R-,61.72.Mm,66.30.Lw,71.15.Mb}

\keywords{Fe grain boundary, H interstitials in metal, kinetic Monte Carlo simulations}

\maketitle

\newcommand{\ee}[1]{\begin{equation} #1 \end{equation}}
\newcommand{\de}[1]{\begin{displaymath} #1 \end{displaymath}}
\newcommand{\br}[0]{{\bf{r}}}
\newcommand{\brp}[0]{{\bf{r^{\prime}}}}
\newcommand{\rp}[0]{{{r^{\prime}}}}
\newcommand{\bR}[0]{{\bf{R}}}
\newcommand{\bG}[0]{{\bf{G}}}
\newcommand{\bv}[0]{{\bf{v}}}
\newcommand{\tp}[0]{t^{\prime}}
\newcommand{\bk}[0]{{\bf{k}}}
\newcommand{\bkQ}[0]{{\bf{k+Q}}}
\newcommand{\ket}[1]{| #1 \rangle}
\newcommand{\bra}[1]{\langle #1 |} 
\newcommand{\braOket}[3]{\langle #1 | #2 | #3 \rangle} 
\newcommand{\braket}[2]{\langle #1 | #2  \rangle} 
\newcommand{\cc}[1]{\begin{center}{\bf{ #1 }}\end{center}}
\newcommand{\ui}[0]{\hat{\bf{i}}}
\newcommand{\ux}[0]{\hat{\bf{x}}}
\newcommand{\uj}[0]{\hat{\bf{j}}}
\newcommand{\uy}[0]{\hat{\bf{y}}}
\newcommand{\uk}[0]{\hat{\bf{k}}}
\newcommand{\uz}[0]{\hat{\bf{z}}}
\newcommand{\ur}[0]{\hat{\bf{r}}}
\newcommand{\e}[1]{{\rm{e}}^{#1}}
\newcommand{\ex}[0]{{\rm{exp}}}
\newcommand{\Gp}[0]{{\bf{G}}_{\parallel}}
\newcommand{\mGp}[0]{G_{\parallel}}
\newcommand{\kp}[0]{{\bf{k}}_{\parallel}}


\section{Introduction  \label{intro}}
Hydrogen is a very common impurity in iron-based materials.  It is incorporated into the material during production and service and exhibits a high mobility within the bulk phase.
It has further been shown that microstructural defects in the material such as vacancies, dislocations, and grain boundaries can trap hydrogen impurities~\cite{Novak10,Ramasubramaniam08,Wen03, Zhong00prb}.  The resulting local accumulation of hydrogen at these defects can then lead to a degradation of the mechanical properties of the material, which is also referred to as hydrogen embrittlement~\cite{Hirth80}.
To explain the complex mechanisms underlying hydrogen embrittlement various approaches have been developed including hydrogen-enhanced local plasticity (HELP)~\cite{Beachem72,BIRNBAUM94,Pezold11}, hydrogen-enhanced decohesion (HEDE)~\cite{ORIANI74, ORIANI77, ORIANI78, Unger88, GERBERICH88} and superabundant vacancy formation~\cite{Nazarov10}.

The most important aspects in trying to understand the behaviour of hydrogen within a material and its role in hydrogen embrittlement are the hydrogen solubility, the interaction of hydrogen with defects, and the hydrogen mobility.
The solubility of hydrogen within pure bcc-Fe is relatively low.  Point and extended defects, however, can provide interstitial sites that are energetically much more favourable for hydrogen than the tetrahedral site in bcc-Fe.  In a recent study we investigated the solution energy of H in the presence of open and close-packed grain boundaries (GB) in bcc- and fcc-Fe~\cite{duprbfeh2011} employing density-functional theory (DFT)~\cite{hohenberg:1964,kohn:1965} calculations.  We find that in general hydrogen prefers the interstitial sites within the grain boundary region (with the exception of close-packed grain boundaries in fcc-Fe).  In particular we find a large energy gain of $0.4-0.5$~eV for hydrogen interstitials at the $\Sigma 5 [001] (310)$ grain boundary in bcc-Fe ($\Sigma 5$ in short) as compared to the bulk region.  This indicates that hydrogen segregates to the grain boundary and is trapped there.

The mobility of hydrogen within pure bcc-Fe bulk is high.  The calculated diffusion barrier for H atoms moving between neighbouring tetrahedral sites is $\sim 0.1$~eV~\cite{Jiang04, duprbfeh2011}.  This is consistent with earlier experimental observations indicating that H  diffuses rapidly within bcc metals~\cite{Fukai85, Nagano1982, Wenzijpf83, Lottner79a, Lottner79b}.
Experimental studies investigating the diffusion of hydrogen in the presence of dislocations, grain boundaries, and phase boundaries show that the measured diffusivity depends on the hydrogen concentration~\cite{kirchheim88}.  In bulk Pd, Pd grain boundaries, and Pd/Al$_2$O$_3$ phase boundaries, the diffusion constant generally increases with increasing H concentration~\cite{kirchheim88}.
Within a theoretical study employing kinetic Monte Carlo (kMC) simulations it was illustrated that the attraction of H to screw dislocations in Fe materials can significantly affect H diffusion~\cite{Ramasubramaniam08}. 
To properly evaluate the mobility of hydrogen within a certain material it is thus important to also consider the effect of the microstructure.

In this study, we employ kMC simulations to address H diffusion in the presence of grain boundaries in bcc-Fe.  In a previous study we found that the H diffusion barriers within the grain boundary region of bcc-Fe are much higher ($0.25 - 0.6$~eV)~\cite{duprbfeh2011} than in the bulk region.  This suggests that  H interstitials diffuse relatively slowly or are effectively immobile within the grain boundary interface, and that therefore the grain boundaries do not provide fast diffusion channels for hydrogen.  Utilising the information about available interstitial sites, diffusion barriers, and solution energies extracted from our DFT calculations, we set up a series of kMC models that represent idealised microstructures.
The first model is an idealised cubic grain structure in bcc-Fe, with and without point defects included in the bulk region of the grain.  Since the model grains are much smaller than grains in the actual material we consider in our second model a parallel arrangement of grain boundary interfaces.  This layered structure naturally introduces an anisotropy in the diffusivity.
The third model presented in this paper represents more detailed the structure of the $\Sigma 5$ grain boundary in bcc-Fe.  Within the above models, H diffusion constants are determined as a function of H concentrations and temperatures. From the results obtained within these model systems general trends for H diffusivity in various structural environments can be extracted.  We then compare the numerical results to the derived analytic expression that describes the macroscopic diffusivity of hydrogen in different microstructures.

The paper is organised as follows. The computational  approach is detailed in Sec.~\ref{meth}.
In Sec.~\ref{cubicgrain} and~\ref{gbpt} the results for H diffusion within the idealised cubic grain structure with and without additional point defects are presented.
The results for the layered structure and for the more detailed model of the  $\Sigma 5$ grain boundary are discussed in Sec.~\ref{layer} and~\ref{sigma5}, respectively.
Our findings are summarised in Sec.~\ref{conls}.


\section{Computational Approach\label{meth}}

We employ kinetic Monte Carlo~\cite{Bortz_1975,Gillespie_1976} simulations to investigate H diffusion under various conditions and within a number of idealised  microstructures. 
Within kMC simulations the time evolution of the system is described by a stochastic trajectory.  The system states along this trajectory are connected by processes associated with a certain probability.  Here, we use a lattice approach, i.e. possible atomic positions are mapped onto a lattice and the system can evolve by atoms (hydrogen) hopping between neighbouring lattice sites (interstitial sites in bcc-Fe bulk and grain boundaries).
Within harmonic transition state theory~\cite{vineyard:1957} the microscopic rate constant for a hop, $k_i$, associated with process $i$ can be written as:
\ee{
\label{rate}
k_{i} = \nu_{0,i}\exp\left(-\Delta E_i/k_{\rm B}T\right) \quad ,
}
where $\nu_{0,i}$ is the attempt frequency, $k_{\rm B}$ is the Boltzmann constant, $T$ is the temperature, and $\Delta E_i$ is the energy barrier associated with process $i$.  For all processes that can occur within our kMC models the corresponding energy barriers were calculated employing DFT calculations (details regarding the DFT calculations can be found in Ref.~\onlinecite{duprbfeh2011}).  For a given system configuration the sum over all  rate constants of all possible processes is evaluated, $k_{\rm tot} = \sum_i k_i$, and a process $p$ to move to the next system state is chosen according to
\ee{
\label{eq:selectp}
\sum_{i=0}^{p-1} k_i < \rho_1 k_{\rm tot} \leq \sum_{i=0}^{p} k_i \quad ,
}
where $\rho_1$ is a uniform random number between 0 and 1.
Since the kMC algorithm simulates a sequence of Poisson processes, the real time evolution for each kMC step can be evaluated as~\cite{Fichthorn_1991}
\ee{
\label{kmctime}
t \to t - \ln(\rho_2)/k_{tot}, 
}
where $\rho_2$ is a second random number between 0 and 1.

The diffusion constant tensor is calculated from the mean square displacement of the hydrogen atoms.  To obtain better statistics on the diffusion constants we follow an approach outlined previously~\cite{kirchheim87, kirchheim88, Ramasubramaniam08}, and divide the kMC trajectory into a number of segments.  The diffusion constant is calculated as the time weighted average of the diffusion constants for each segment $i$
\ee{
\label{df}
D_{kk} = \sum_{i} D_{kk, i}  \Delta t_{i} /t \quad ,
}
where for the tetragonal structures analysed in this paper  the diagonal components of the diffusion tensor $D_{kk}$  suffice; $\Delta t_i = t_i - t_{i-1}$ is the time length of segment $i$, $t$ is the total length of the kMC trajectory, and
\ee{
\label{dfseg}
D_{kk, i} =
\langle [r_{k}(t_{i}) - r_{k}(t_{i-1})]^{2}\rangle /2 \Delta t_{i}
}
is the diffusion constant for segment $i$.
Here, $r_k(t_i)$ is the position of a H atom in $k$-direction (with $k=x,y,z$) at time $t_i$ and $\langle \dots \rangle$ denotes the average over all particles.  The overall diffusion constant $D$ is is defined as the average of the diagonal components of the diffusion tensor, $D = \frac{1}{3} \sum_{k} D_{kk}$.

Similarly, we define the probability $p_s$ of finding hydrogen atoms in certain sites $s$ (e.g. bulk, interface or point defect sites) as
\ee{
\label{runavep}
p_{s} = \frac{\sum_{i} p_{i,s} \Delta t_{i}}{t} \quad ,
}
with
\ee{
\label{instp}
p_{i,s} = \frac{N_{i,{\rm H},s}}{N_{\rm H}} \quad ,
} 
where $N_{\rm H}$ is the total number of hydrogen atoms in the simulation and $N_{i,{\rm H},s}$ is the number of hydrogen atoms at site type $s$ in the trajectory segment $i$.

The analytic expression for the diffusion constant $D$ within a symmetric lattice and in the limit of low hydrogen concentrations is given by the classical Arrhenius expression
\ee{
\label{simpledc} 
D = D_0 \exp\left(-\Delta E/ k_{\rm B}T\right) \quad ,  
}
with $D_0 = \Gamma a_0^2 \nu_0$. $\Delta E$ is the diffusion barrier, $a_{0}$ is the jump distance for a diffusion hop, and $\Gamma$ is a geometric prefactor that is related to the connectivity of each site to its neighbouring sites.  For isotropic diffusion the  geometric prefactor follows~\cite{diffconst:1993}:
\ee{
\label{geoprefactor} 
\Gamma = \frac{n}{2 d} \quad , 
}
where $n$ is the number of equivalent nearest neighbour sites and $d$ is the dimensionality of the system.

To obtain an analytic expression for the probability to find hydrogen in various types of interstitial sites we assume a grand canonical ensemble and express the total number of H atoms within the system as~\cite{kirchheim88}:
\ee{
\label{Eq:genHnum}
N_{\rm H} = \sum_{s} \frac{n_s}{1 + \exp\left((E_{s} -\mu)/ k_{\rm B}T\right)} \quad ,
}
where the sum runs over the various interstitial site types $s$, $n_s$ is the number of sites of type $s$, $E_{s}$ is the solution energy of hydrogen at site type $s$, and $\mu$ is the chemical potential.
The probability to find hydrogen at  site type $s$ is therefore 
\ee{
\label{Eq:genprbty}
p_{s} = \frac{N_{\text{H},s}}{N_{\text{H}}} = \frac{1}{N_{\text{H}}} \cdot \frac{n_{s}}{1 + \exp\left((E_{s}-\mu)/ k_{\rm B}T\right) }
           \quad.
}
The analytic results for the probability of finding hydrogen in certain sites can directly be compared with our numerical results employing Eq.~(\ref{runavep}).


\section{Idealised Cubic Grain\label{cubicgrain}}

\begin{figure}
\includegraphics[width=0.4\textwidth]{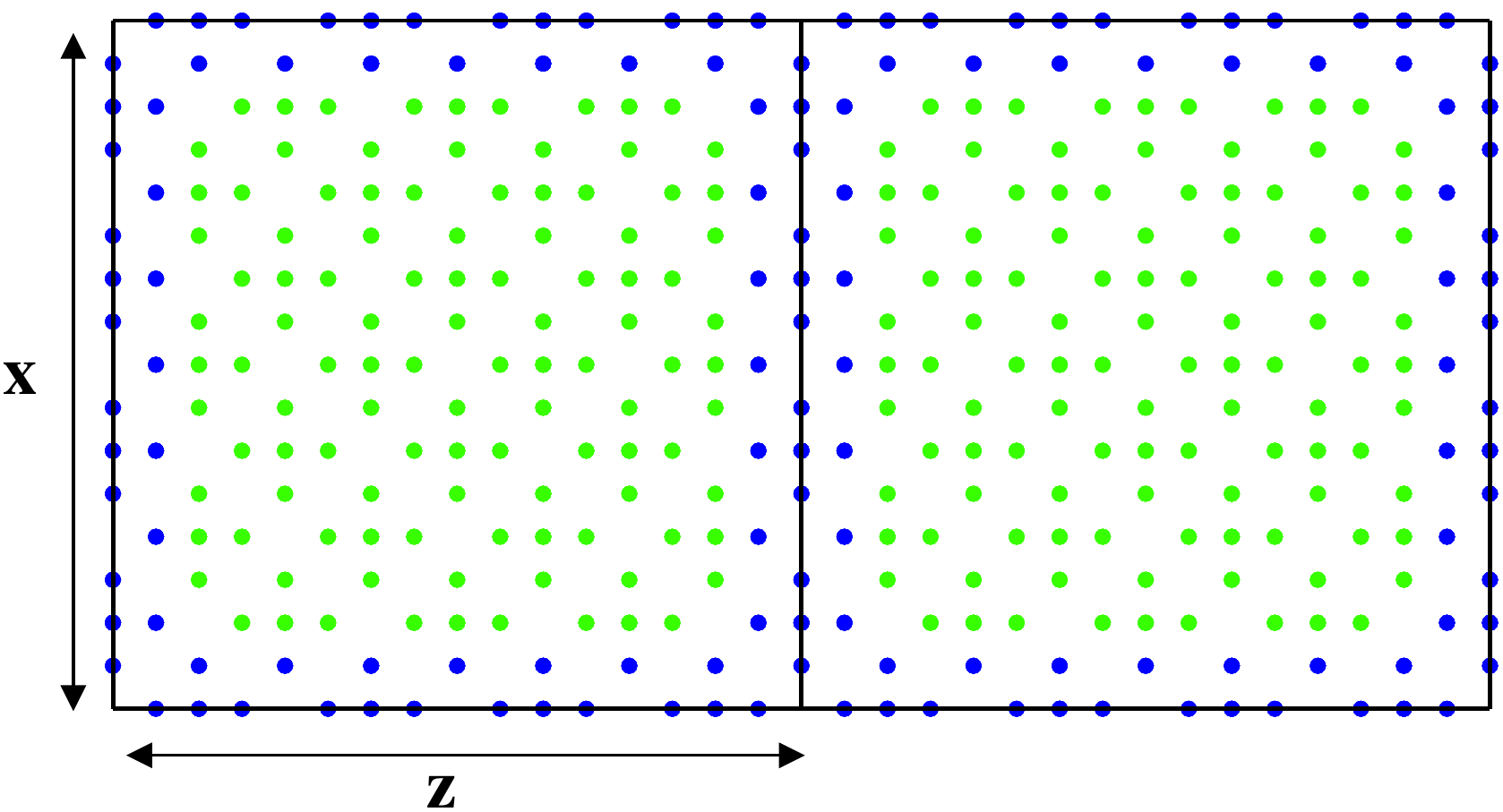}
\caption{(Colour online) Model of an idealised cubic grain in bcc-Fe. The spheres represent tetrahedral interstitial sites in bcc-Fe, each having four nearest neighbours at the same distance.  Interface sites are shown in blue, bulk sites are shown in green.
\label{fig:cubicgrain}}
\end{figure}

\begin{table}
\caption{Diffusion barriers, $\Delta E_i$, between various sites $i$ within the idealised cubic grain.
 \label{tab:hoppingrate}}
\begin{ruledtabular}
\begin{tabular}{lc} 
process $i$   & $\Delta E_{i}$ (eV) \\ \hline
bulk site $\rightarrow$ bulk site       &   $0.088$ \\
bulk site $\rightarrow$ interface site      &   $0.088$   \\
interface site  $\rightarrow$ bulk site      &   $0.600$  \\
interface site  $\rightarrow$ interface site     &   $0.250$    \\
bulk site  $\rightarrow$ point defect      &   $0.088$    \\
point defect  $\rightarrow$ bulk site     &   $0.750$    \\
\end{tabular}
\end{ruledtabular}
\end{table}

To study hydrogen diffusion within an idealised cubic grain structure of bcc-Fe we use the model shown in Fig.~\ref{fig:cubicgrain}. In this simple model we only consider two different site types, bulk and interfaces sites, shown as blue and green spheres in Fig.~\ref{fig:cubicgrain}, respectively.   All sites in the kMC model are arranged as tetrahedral sites in bcc-Fe (preferred interstitial site for hydrogen), each having four nearest neighbours at the same distance.   Thus the two site types have the same geometry, but different solution energies and respective diffusion barriers.  The grain boundaries between bulk regions are represented by three layers of interface sites. 
Within the kMC simulations we use a simulation cell with a side length of  $x = 45.312$~\AA\ containing 5952 interface and 43200 bulk sites.  The simulation cell is periodically repeated in all three dimensions.

The total number of H atoms within the system is 
\ee{
\label{nofh1}
N_{\rm H} = N_{\rm H, intf} +  N_{\rm H, bulk} \quad ,
}
where $N_{\rm H, intf}$ and  $N_{\rm H, bulk}$ are the number of H atoms at the interface and bulk sites, respectively. Correspondingly, we express the number of H atoms within the two site types as
\ee{
\label{gcenhtrap}
N_{\rm H, intf} = \frac{n_{\rm intf}}{1 + \exp\left((E_{\rm intf}-\mu)/ k_{B}T\right)} \quad ,
}
and
\ee{
\label{gcenhbulk}
N_{\rm H, bulk} =  \frac{n_{\rm bulk}}{1 + \exp\left((E_{\rm bulk}-\mu)/ k_{B}T\right)} \quad .
}
The probability of finding H at bulk sites is therefore
\ee{
p_{\rm bulk} = \frac{N_{\rm H, bulk}}{N_{\rm H, bulk} + N_{\rm H, intf}}  \quad .
\label{gceprobty}
} 

To determine the solution energies and respective diffusion barriers between the two site types we use the results of our recent DFT study~\cite{duprbfeh2011}.  The bulk sites in our model correspond to tetrahedral interstitial sites in bcc-Fe, and the interface sites correspond to the most stable interstitial site within the grain boundary region of the $\Sigma 5$~GB.
According to our DFT calculations the interstitial site at the $\Sigma 5$~GB provides a trap for H interstitials with a binding energy of $\Delta E_{\rm bind} = E_{\rm bulk} - E_{\rm intf} = 0.512$~eV.  There are four different processes for hydrogen diffusion between bulk and interface sites:  bulk $\rightarrow$ bulk, bulk $\rightarrow$ interface, interface $\rightarrow$ bulk, and interface $\rightarrow$ interface.  The energy barriers, $\Delta E_i$, to determine the rate constants for these processes according to Eq.~(\ref{rate}) are likewise taken from our recent DFT study~\cite{duprbfeh2011} and are listed in Table~\ref{tab:hoppingrate}.  The attempt frequency $\nu_0$ is set to  $10^{13}$~s$^{-1}$ for all processes.

\begin{figure}
\includegraphics[width=0.36\textwidth]{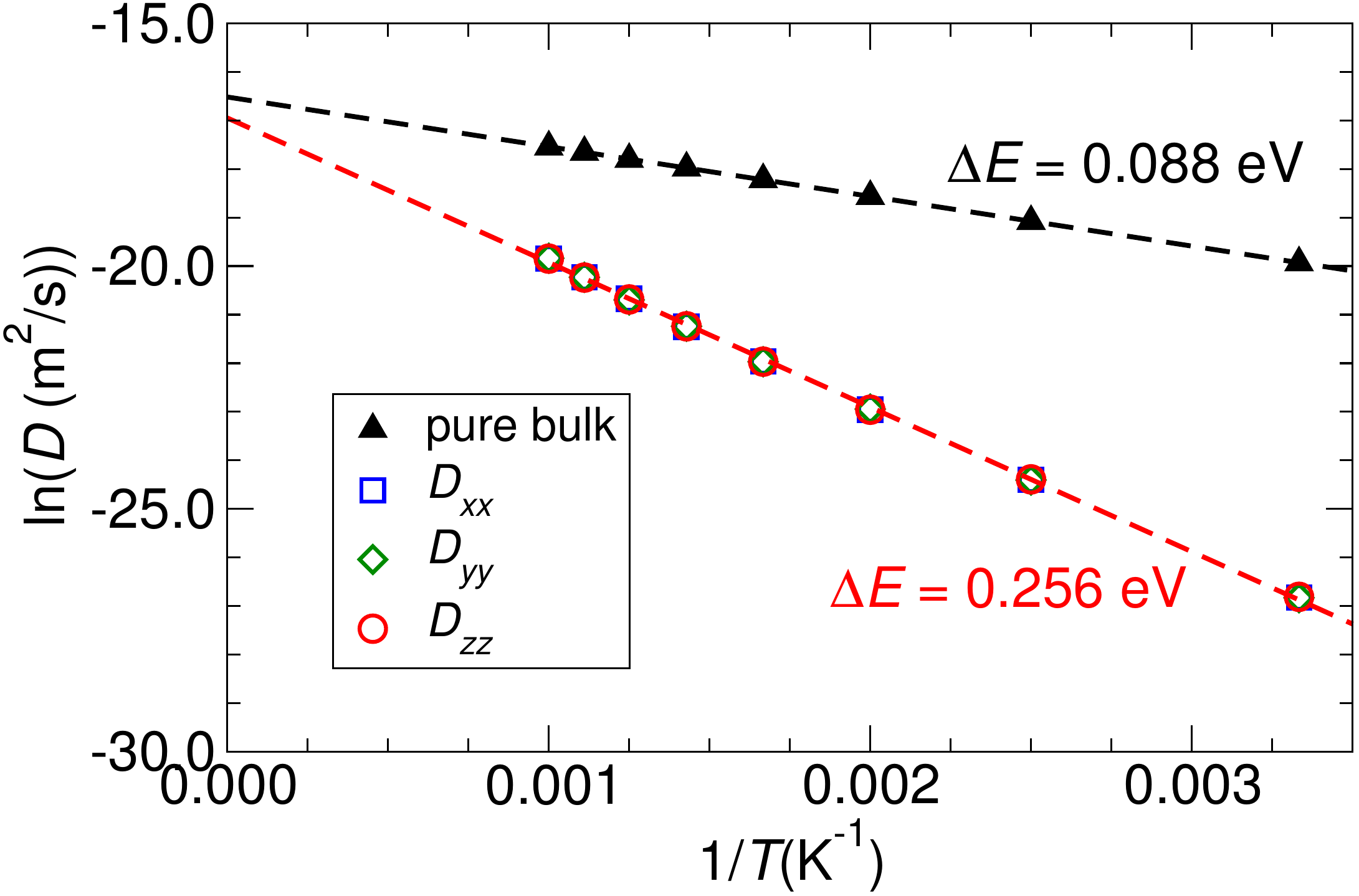}
\caption{(Color online) Logarithm of the diffusion constants for hydrogen as a function of the inverse temperature within the dilute limit.  The black triangles are the results for hydrogen diffusion between tetrahedral sites in perfect bcc-Fe. The blue, green, and red symbols represent the diagonal components of the diffusion tensor for the idealised cubic grain structure confirming an isotropic diffusion of hydrogen.  The dashed lines correspond to linear fits of the kMC data.
\label{fig:Cell444-df}}
\end{figure}

In Fig.~\ref{fig:Cell444-df} results for hydrogen diffusion constants for the cubic grain structure and for a perfect bulk structure are compared within the dilute limit.  The perfect bulk structure contains only bulk sites with a diffusion barrier of $0.088$~eV between neighbouring sites and serves as a reference within our simulations.  In Fig.~\ref{fig:Cell444-df} the logarithm of the diffusion constants is shown as a function of the inverse temperature.
If we assume an Arrhenius-like behaviour, cf. Eq.~(\ref{simpledc}), then the slope corresponds to $-\Delta E/k_{\rm B}$ and the $y$-intercept to $\ln(D_0)$.  For our reference, perfect bulk system a barrier of $\Delta E = 0.088$~eV is extracted from the slope. We obtain $\ln(D_0) = -16.52$, as expected from the theoretical value of $\ln(\Gamma a_0^2 \nu_0)$ with $\Gamma = 4/6$, $a_0 = 1.001$~\AA, and $\nu_0 = 10^{13}$. 
Within the grain structure the three diagonal components of the diffusion tensor, $D_{xx}$, $D_{yy}$, and $D_{zz}$, are equivalent confirming the isotropic diffusion of hydrogen.  
The slope of the linear fit to the simulation data yields a value for the effective diffusion barrier of $\Delta E = 0.256$~eV, which corresponds to the barrier for an interface $\to$ interface hop, suggesting that diffusion mainly takes place within the interface region.  The fitted value of $\ln(D_0) = -16.95$ is smaller than in perfect bulk.
Assuming diffusion only within the interface region there are only two adjacent interface sites, i.e. $\Gamma = 2/6$, yielding a theoretical value of $\ln(D_0) = -17.21$.
It can also be seen that within the dilute limit diffusion is slower in the grain structure than in the perfect bulk structure.  
This can be explained by the fact that  within the grain structure hydrogen is confined to the interface region which exhibits a higher barrier for diffusion.

\begin{figure}
\includegraphics[width=0.44\textwidth]{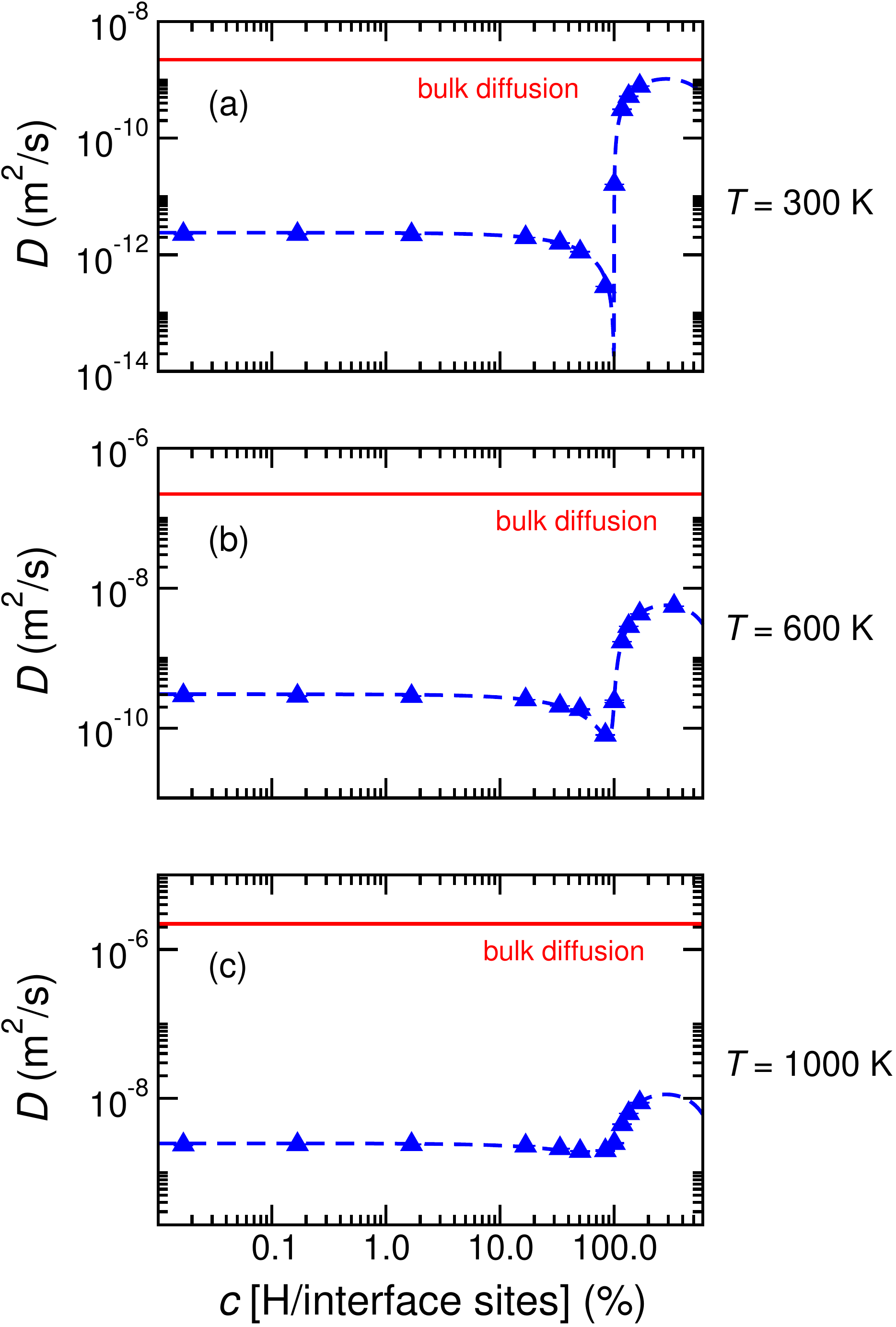}
\caption{(Color online) Diffusion constants as a function of H concentrations in an idealised cubic grain for temperatures of (a) 300~K, (b) 600~K, and (c) 1000~K.  The symbols denote numerical results from the kMC simulations, the dashed lines are obtained from  Eq.~(\ref{eq:diffgeneral}), respectively.  The red curve represents the diffusivity of  H  in perfect bcc-Fe within the dilute limit.
\label{fig:Cell444-com}}
\end{figure}

In Fig.~\ref{fig:Cell444-com} hydrogen diffusion constants as a function  of H concentration are shown for temperatures of $T = 300$, 600, and  1000~K, respectively.  The kMC results are illustrated by blue triangles.  
For all three temperatures the diffusivity is constant for low H concentrations, it then decreases as the number of H atoms approaches the number of interface sites, before it again  increases quickly to a considerably higher value.  This behaviour can be explained as follows:  (I)  at low H concentrations diffusion mainly takes place in the interface region and the associated diffusivity is lower than in perfect bulk (red line in Fig.~\ref{fig:Cell444-com}) due to a higher diffusion barrier in the interface region; 
(II) when the H concentration approaches the number of interface sites, a dip occurs in the  diffusion constant.  This is due to  blocking in the interface region,  i.e. hydrogen is mainly confined to the interface region filling all the interface sites and effectively blocking diffusion processes;
(III) at high H concentrations ($N_{\rm H} > n_{\rm intf}$), the overall diffusivity results mainly from bulk diffusion, which is  much faster and thus the diffusivity increases.    
The dip in the diffusivity for hydrogen concentrations close to the number of interface sites is less pronounced for higher temperatures, because at higher temperatures H atoms have a higher probability to escape from the interface to the bulk region.
At very high hydrogen concentrations the diffusivity decreases again due to blocking in the bulk region.

Our numerical results from the kMC simulations can also be described by an analytic expression.   The overall diffusion constant is a combination of diffusion in the interface and in the bulk region and can thus be approximated as the weighted sum over the two contributions:  
\ee{
\label{eq:diffgeneral}
D(T,N_{\rm H})  =  \sum_s p_s(T) D_{0,s} \exp\left(\frac{-\Delta E_s}{k_{{\rm B}}T}\right) \Xi_s(T,N_{\rm H}) \quad .
}
The sum is weighted by the probabilities, $p_s$, to be in the bulk or interface region, cf. Eq.~(\ref{Eq:genprbty}).  In addition the blocking~\cite{Kirchheim1jncs85} in the two regions is accounted for by the factor 
\ee{
\label{eq:blocking}
\Xi_s(T,N_{\rm H}) = \left(1 - \frac{\overline{N}_{{\rm H},s}(T,N_{\rm H})}{n_{s}}\right) \quad ,
}
where $\overline{N}_s(T,N_{\rm H}) =  p_s(T) N_{\rm H}$ is the average number of hydrogen atoms at site type $s$ and $n_s$ is the number of sites of type $s$.  Diffusion in between the two site types is assumed to be in equilibrium and these processes do not significantly contribute to the overall diffusivity.  
The prefactor $D_{0,s}$ contains the geometric prefactor $\Gamma_s$, the attempt frequency $\nu_0$ and the hopping distance $a_0$, with $\nu_0 = 10^{13}$~s$^{-1}$ and $a_0 = 1.001$~\AA.  
The geometric prefactors for the bulk and interface regions, $\Gamma_{\rm bulk}$ and $\Gamma_{\rm intf}$ are obtained by fitting to the numerical kMC data.

Before comparing the analytic and numerical results of the diffusion constants we verify that the hydrogen concentration in the two regions is indeed in equilibrium.  For this, we evaluate the probability to be in the bulk region, $p_{\rm bulk}$, within our kMC simulation according to Eq.~(\ref{runavep}) and compare it to the analytic result in Eq.~(\ref{Eq:genprbty}).
\begin{figure}
\includegraphics[width=0.34\textwidth]{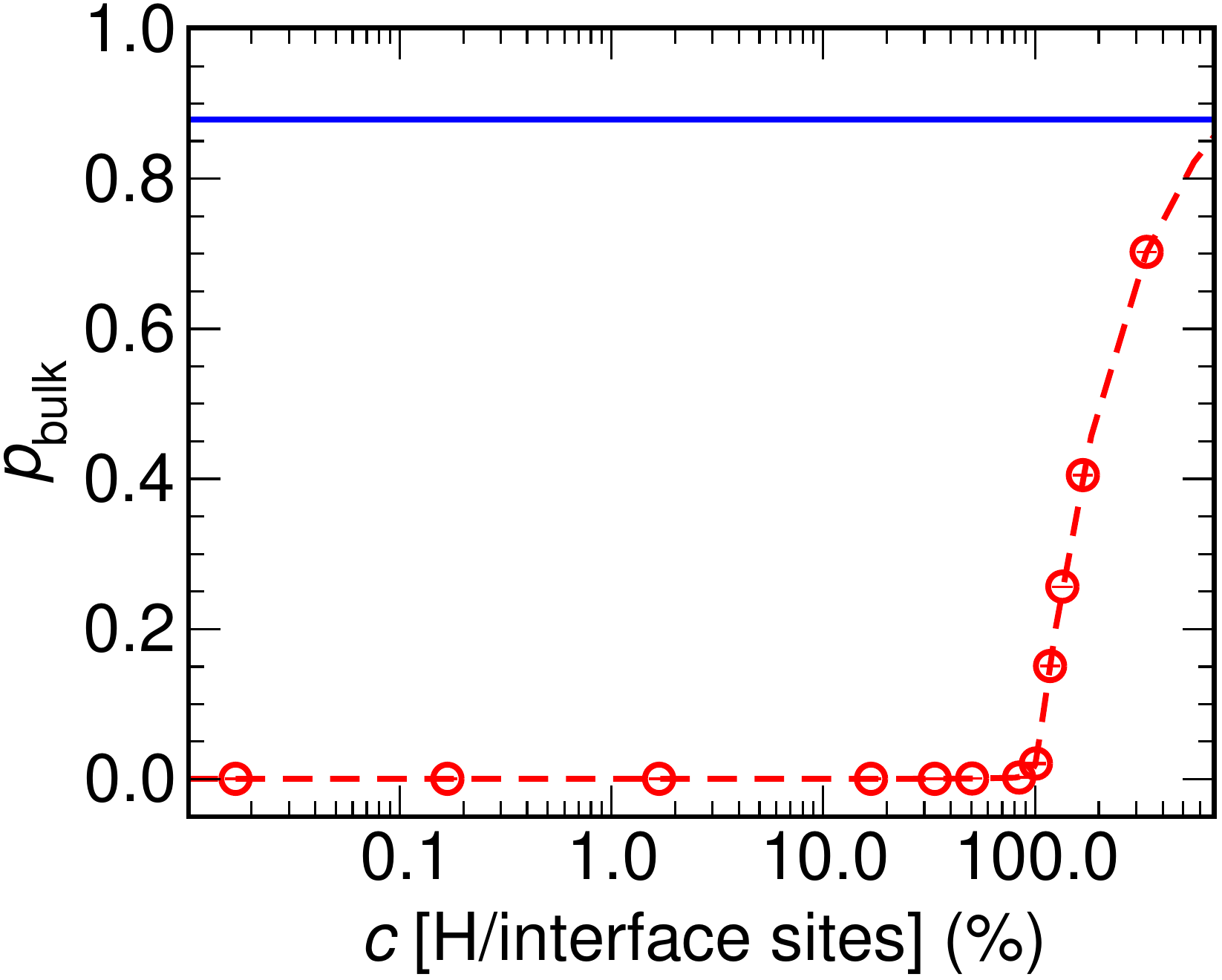}
\caption{(Color online) Probability of finding hydrogen in the bulk region of an idealised cubic grain  at $T = 600$~K as a function of H concentration.  The circles represent numerical results from the kMC simulations, the dashed line shows the result from the grand canonical ensemble model, cf. Eq.~(\ref{Eq:genprbty}).  The solid blue line marks the theoretical limit $n_{\rm bulk}/(n_{\rm bulk} + n_{\rm intf})$.
\label{fig:cell-1000K-gce-p}}
\end{figure}
In Fig.~\ref{fig:cell-1000K-gce-p} the results are shown for a temperature of $T = 600$~K.  The analytic result agrees well with the kMC results. For high H concentrations, $p_{\rm bulk}$ converges to the theoretical limit $n_{\rm bulk}/(n_{\rm bulk} + n_{\rm intf})$. 
Comparing Figs.~\ref{fig:Cell444-com} and~\ref{fig:cell-1000K-gce-p} indicates that for a wide range of concentrations, hydrogen is indeed confined to the interface region, even at $T = 600$~K. Only when the interface sites are filled with hydrogen diffusion in the bulk region enhances the overall diffusivity.

Using our kMC data at $T = 300$, 600, and  1000~K, the fitted geometric prefactors are  $\Gamma_{\rm intf} = 0.377$ and  $\Gamma_{\rm bulk} = 0.644$.  As shown in Fig.~\ref{fig:Cell444-com} the analytic results are in excellent agreement with the kMC results.  The fitted values for the geometric prefactors also agree well with their theoretical values.  In the bulk region each site has four nearest neighbours yielding $\Gamma_{\rm bulk} = 4/6 = 0.667$, whereas the interface region is a 3D network of 2D grain boundary plains with only two nearest neighbours for each interface site, yielding $\Gamma_{\rm intf} = 2/6 = 0.333$.
The small deviation of the numerical fitted values from the ideal theoretical ones is mainly due to diffusion processes in the vicinity of neighbouring bulk and interface sites where the local connectivity differs from the ideal one.


\section{Idealised Cubic Grain with Isolated Point Defects\label{gbpt}}

In addition to the interface and bulk sites we next included point defects in the bulk region as a third site type into our model.  Point defects such as vacancies or substitutional atoms are typically present in materials and may as well influence the diffusion  of hydrogen.  Our idealised cubic grain model with point defects contains 400 point defects within the bulk region, 5952 interface sites and 42800 bulk sites.  The point defect sites have the same geometry as bulk and interface sites, but different energetics.  Since hydrogen binds even stronger to vacancies than to interface sites~\cite{Tateyama03prb}, the point defect sites are also considered to be more stable in our kMC model.  The corresponding microscopic diffusion barriers are summarised in Table~\ref{tab:hoppingrate}.

Since the point defects are not connected to each other there is no direct hop from one point defect to the next.  Hence, the geometric prefactor, $\Gamma_{\rm trap}$ for point defects in Eq.~(\ref{eq:diffgeneral}) is expected to be zero.  However, the point defects alter the distribution of hydrogen within the system and thus the corresponding probabilities, $p_s$.

\begin{figure}
\includegraphics[width=0.34\textwidth]{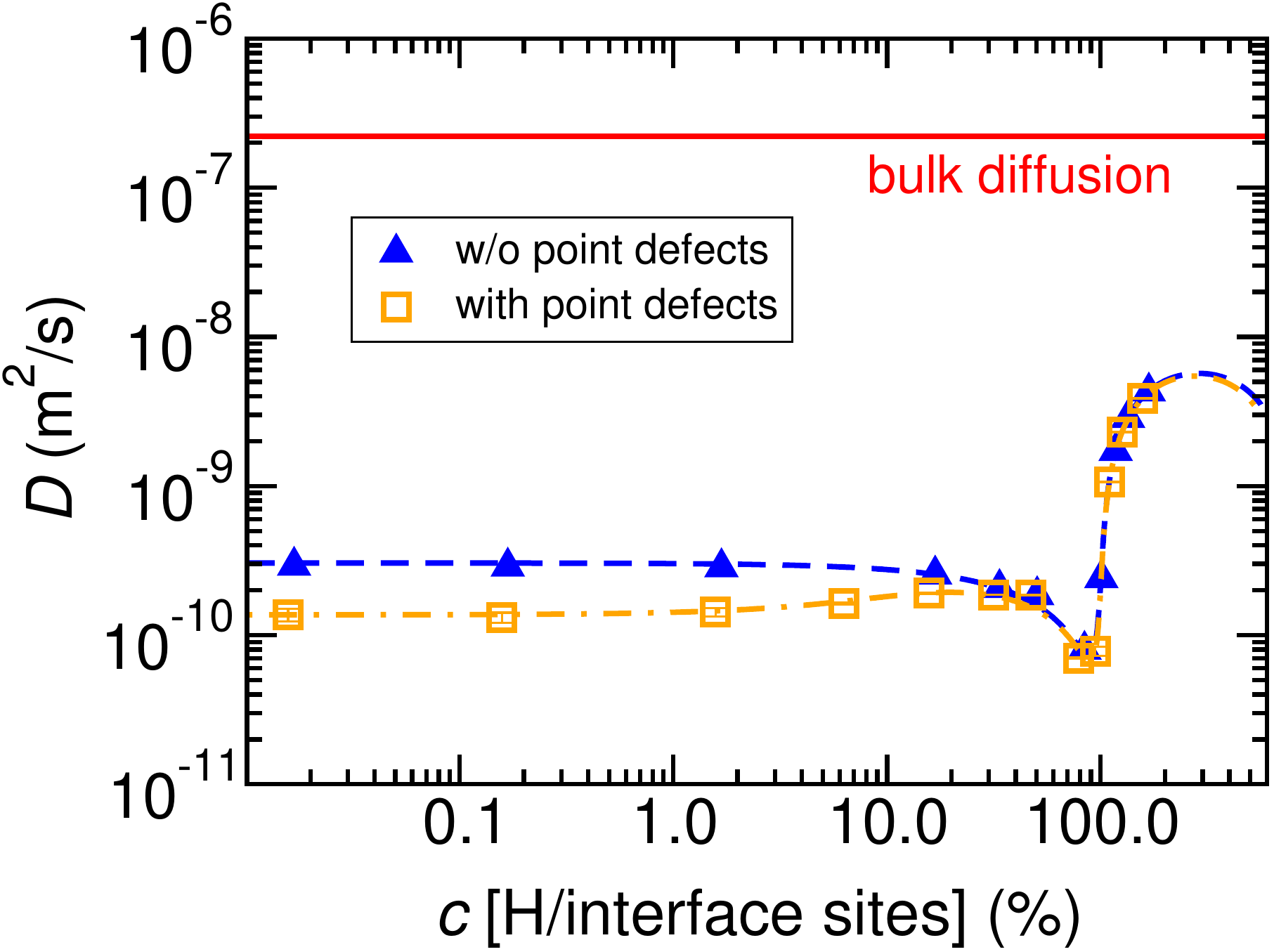}
\caption{(Color online) Diffusion constants as a function of H concentration at $T = 600$~K in an idealised cubic grain (blue (dark) triangles) and in and idealised grain with additional 400 point defects in the bulk region (orange (light), squares).  The symbols denote results from the kMC simulations, dashed/dotted lines indicate analytic results according to Eq.~(\ref{eq:diffgeneral}).  The red curve represents the diffusivity of  H  in  perfect bcc-Fe within the dilute limit.
\label{fig:Cell444-pt-dfvc600}}
\end{figure}

In Fig.~\ref{fig:Cell444-pt-dfvc600} the results for H diffusivities in an idealised cubic grain with and without point defects are compared for a temperature of $T = 600$~K.  At low concentrations most hydrogen atoms are  trapped at the point defects and hence do not directly contribute to the overall diffusivity. Therefore, the diffusion constants are lower as compared to those in the same grain without point defects.  At higher H concentrations the effect becomes less pronounced and eventually all point defects and interface sites are filled, and hydrogen atoms in the bulk region dominate the overall diffusion.

For the grain model including point defects the fitted geometric prefactors  for the interface region, bulk region and point defects are  $\Gamma_{\rm intf} = 0.376$, $\Gamma_{\rm bulk} = 0.634$, and $\Gamma_{\rm trap} = 0.0$, respectively.  The analytical results obtained from Eq.~(\ref{eq:diffgeneral}) are in excellent agreement  with the kMC results as shown in Fig.~\ref{fig:Cell444-pt-dfvc600}. Note, that the fitted geometric prefactors for the model without point defects are  $\Gamma_{\rm intf} = 0.377$ and $\Gamma_{\rm bulk} = 0.644$.  The  geometric prefactor for the  bulk region decreases slightly in the presence of point defects, since the point defects reduce the effective number of equivalent nearest-neighbour sites within the bulk region.

In the limit of low hydrogen concentrations and low temperatures Eq.~(\ref{eq:diffgeneral}) reduces to an expression previously used to describe H diffusion in the presence of point traps in iron materials~\cite{Oriani70, Ramasubramaniam08}.  Assuming there are only bulk and isolated point defect trapping sites with $\Gamma_{\rm trap} = 0.0$, Eq.~(\ref{eq:diffgeneral}) reads:
\begin{eqnarray}
\label{eq:difftrap}
\lefteqn{D(T,N_{\rm H}) =}  \\ \nonumber
               &   &  p_{\rm bulk} D_{0,{\rm bulk}} \exp\left(\frac{-\Delta E_{\rm bulk}}{k_{\rm B}T}\right) \Xi_{\rm bulk}(T,N_{\rm H}) \quad ,
\end{eqnarray}
with 
\begin{eqnarray}
\label{layerprbty}
\lefteqn{p_{\rm bulk} =}  \\ \nonumber 
           & &  \frac{\frac{n_{\rm bulk}}{1 + \exp\left((E_{\rm bulk}-\mu)/ k_{\rm B}T\right)}}  
           {\frac{n_{\rm trap}}{1 + \exp\left((E_{\rm trap}-\mu)/ k_{\rm B}T\right)} 
           + \frac{n_{bulk}}{1 + \exp\left((E_{\rm bulk} -\mu)/ k_{\rm B}T\right)}}.
\end{eqnarray}
For both low H concentrations and low temperatures we can approximate 
$\exp(E_i-\mu/ k_{\rm B}T) \gg 1$ 
and
\begin{eqnarray}
\label{layerprbty2} 
p_{\rm bulk} &\approx&  \frac{\frac{n_{\rm bulk}}{\exp\left((E_{\rm bulk}-\mu)/ k_{\rm B}T\right)}}  
           {\frac{n_{\rm trap}}{\exp\left((E_{\rm trap}-\mu)/ k_{\rm B}T\right)} + \frac{n_{\rm bulk}}{\exp\left((E_{\rm bulk} -\mu)/ k_{\rm B}T\right)}} \nonumber \\
     &  = & \frac{1}  
           {1 + \frac{n_{\rm trap}}{n_{\rm bulk}} \exp\left(\frac{\Delta E_{\rm bind}}{k_{\rm B}T}\right)}  \quad , 
\end{eqnarray}
where $\Delta E_{\rm bind} = E_{\rm bulk} - E_{\rm trap}$ is the binding energy of hydrogen to the trapping site.
Furthermore, for low H concentrations $\Xi_{\rm bulk} \to 1$ and substituting Eq.~(\ref{layerprbty2}) into Eq.~(\ref{eq:difftrap}) leads to the known expression~\cite{Oriani70, Ramasubramaniam08}
\begin{small}
\begin{eqnarray}
\label{pointdcfinal}
\lefteqn{D(T)  = }  \\ \nonumber
        & & D_{0,{\rm bulk}} \exp\left(\frac{-\Delta E_{\rm bulk}}{k_{{\rm B}}T}\right) \left(1 + \exp\left(\frac{\Delta E_{\rm bind}}{k_{\rm B}T}\right) \frac{n_{\rm trap}}{n_{\rm bulk}}\right)^{-1}, 
\end{eqnarray}
\end{small}
As can be seen from Eq.~(\ref{pointdcfinal}) the effect of point defects becomes negligible for small numbers of trapping sites, a small binding energy or high temperatures.


\section{Idealised layered structure\label{layer}}

\begin{figure}
\includegraphics[width=0.45\textwidth, angle=0]{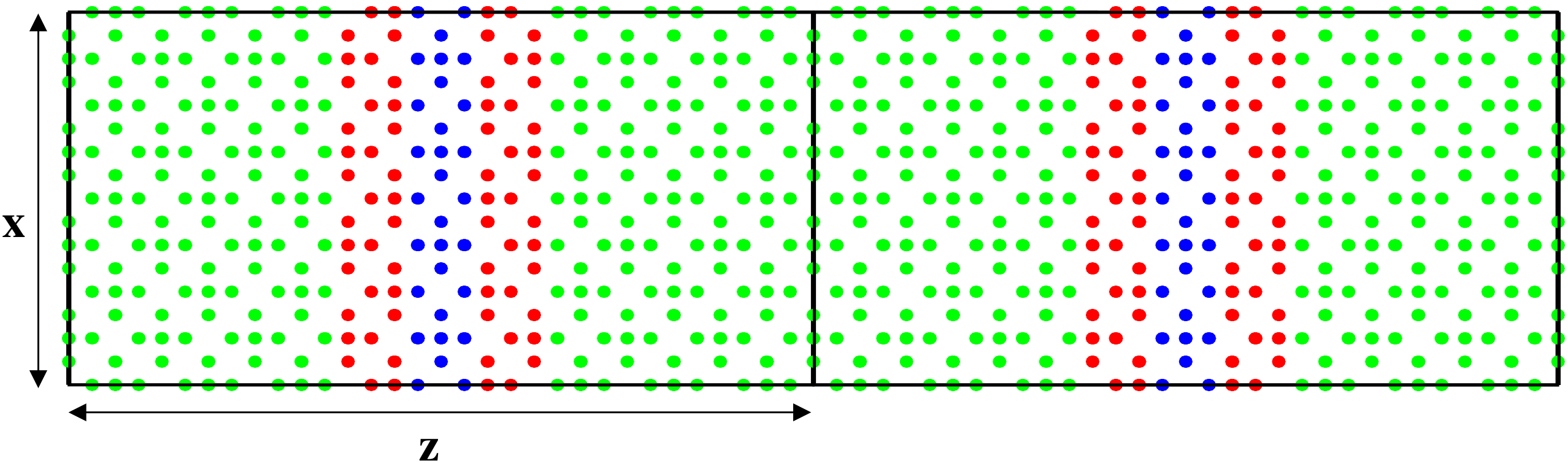}
\caption{(Colour online) Model of an idealised layered structure of
  parallel grain boundary regions in bcc-Fe.  The spheres represent
  tetrahedral interstitial sites in bcc-Fe, each having four nearest
  neighbours at the same distance.  Interface sites, intermediate sites,
  and bulk sites  are shown in blue, red, and green, respectively.
  \label{fig:layergbmodel}}
\end{figure}

The idealised cubic grains discussed in Sections~\ref{cubicgrain} and~\ref{gbpt} are  small compared to typical grain sizes in materials.  In atomistic simulations grain boundary structures are thus often modelled as parallel arrangements of interface planes to describe their 2-dimensional nature.
As a second microstructure  model we investigate the diffusion of hydrogen within such an idealised layered structure as shown in Fig.~\ref{fig:layergbmodel}. 

All sites in the kMC model reflect the geometry of tetrahedral interstitial sites in bcc-Fe.  The GB region is represented by three layers of interface sites indicated by blue spheres in Fig.~\ref{fig:layergbmodel}.
Additionally, the three layers above and below the interface region are marked as intermediate sites (red spheres).
The structure is highly anisotropic, and  diffusion within the interface layers and perpendicular to them is expected to differ significantly.  Within the kMC simulations cells with side lengths of  $x = y = 11.328$~\AA, and  $z = n \times 11.328 $~\AA\ (with $n = 1 - 5$) are used that are periodically repeated in all three dimensions.  

\begin{figure}
\includegraphics[width=0.36\textwidth]{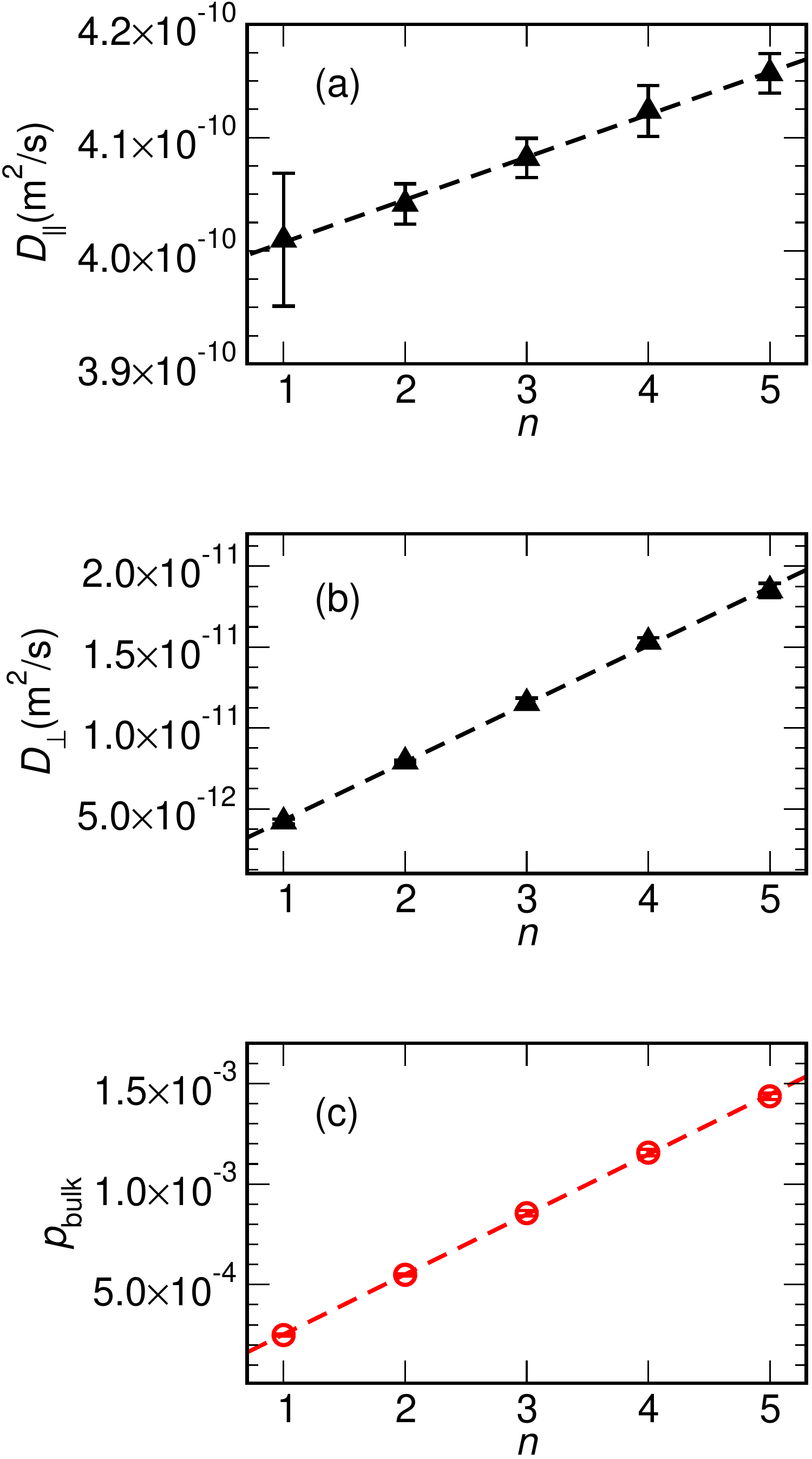}
\caption{(Colour online) H diffusivity and H distribution as a function of the interface-interface distances within the layered structure: (a) components of the diffusion tensor parallel to the interface, $D_{\parallel} = 1/2 (D_{xx} + D_{yy})$; (b) component of the diffusion tensor perpendicular to the interface,  $D_{\perp} = D_{zz}$;  (c) probability of finding H in the  bulk region, $p_{\rm bulk}$.  Symbols denote results from the kMC simulations, dashed lines correspond to linear fits.
\label{fig:layerdf}}
\end{figure}

In a first step we investigate the diffusivity of H as a function of the interface-interface distance.  Here, the intermediate sites are equivalent to bulk sites, the corresponding barriers are taken from case~(I) in Table~\ref{tab:3barrier}.
The results of the kMC simulations  at $T = 600$~K are shown in  Fig.~\ref{fig:layerdf}. 
In Fig~\ref{fig:layerdf}~(a) and (b) the diagonal components of the diffusion tensor parallel, $D_{\parallel} = 1/2 (D_{xx} + D_{yy})$, and perpendicular, $D_{\perp} = D_{zz}$, to the interface layers are depicted, respectively.  Diffusion perpendicular to the interface layers is much slower than parallel to the interface, but in both directions the diffusivity exhibits a linear dependence on the distance between the interface layers.  
This is due to the probability of finding H atoms within the bulk region, $p_{\rm bulk}$.  According to Eq.~(\ref{eq:diffgeneral}) at low H concentration (with $\Xi_{\text{bulk}} \approx \Xi_{\text{intf}} \approx 1$) the overall diffusion constant depends linearly on $p_{\rm bulk}$ (with $p_{\rm intf} = 1 - p_{\rm bulk}$) where the slope depends on the difference in diffusivity in the bulk and interface region.  Since the diffusivity in the bulk region is  higher than in the interface region the overall diffusivity increases linearly with increasing $p_{\rm bulk}$.
$p_{\rm bulk}$ again increases likewise linearly with increasing interface-interface distance as shown in Fig.~\ref{fig:layerdf}~(c).  Since within our current model the condition 
$\exp(\Delta E_{\rm bind}/ k_{\rm B}T) \cdot n_{\rm intf}/n_{\rm bulk} \gg 1$ 
holds (i.e. there is a notable number of interface sites with a significant binding energy), Eq.~(\ref{layerprbty2}) can further be simplified
\ee{
\label{layerprbty3}
p_{\rm bulk} =   \frac{n_{\rm bulk}}{n_{\rm intf}} \exp\left(\frac{-\Delta E_{\rm bind}}{k_{\rm B}T}\right) \quad ,
}  
suggesting that $p_{\rm bulk}$ is proportional to the ratio of bulk and interface sites, $n_{\rm bulk}/n_{\rm intf}$, at a given  temperature $T$. As $n_{\rm bulk}$ increases linearly with the interface-interface distance, whereas $n_{\rm intf}$ remains constant, their ratio increases linearly and so does $p_{\rm bulk}$.

Diffusion perpendicular to the interface layers can be described in an even simpler model.  Since the interface region provides traps for H atoms with an energy gain of 0.512~eV  and H diffuses rapidly within the bulk region, the diffusion perpendicular to the interface can be approximated by  H hopping between two adjacent grain boundary planes.  Within this 1D-model the hoping rate is  $k_{\perp} = \nu_{\perp} \e{- \Delta E_{\perp}/{\rm k_{B}} T}$, and 
\ee{
\label{eq:1ddf}
D_{\perp} = \Gamma_{\perp} a^2_{\perp} \nu_{\perp}  \exp\left(\frac{-\Delta E_{\perp}}{{\rm k_{B}} T}\right) \quad .
}
Here, $\Gamma_{\perp} = 2/2 = 1$ and $a_{\perp}$ is the interlayer distance.  As discussed, the probability of finding H in the  bulk region, $p_{\rm bulk}$ is proportional to the interlayer distance and thus is the time, $\tau$, that H spends in the bulk region, yielding the relation $p_{\rm bulk} \sim a_{\perp} \sim \tau = \frac{1}{\nu_{\perp}}$.  Hence, the  attempt frequency  $\nu_{\perp}$ is inversely proportional to $a_{\perp}$, and  as a result, the diffusion constant $D_{\perp}$ is  proportional to $a_{\perp}$, which is consistent with the results shown in Fig.~\ref{fig:layerdf}~(b).  
The linear dependence of $1/\nu_{\perp}$ on the interlayer distance is strictly only valid if the  bulk region is much thicker than the interface region. With increasing interface-interface distance and in the limit of an infinitely thin interface layer the effective hopping barrier $\Delta E_{\perp}$ converges to the barrier for escaping from the interface region $\Delta E_{\text{intf} \to \text{bulk}}$.

\begin{figure}
\includegraphics[width=0.37\textwidth]{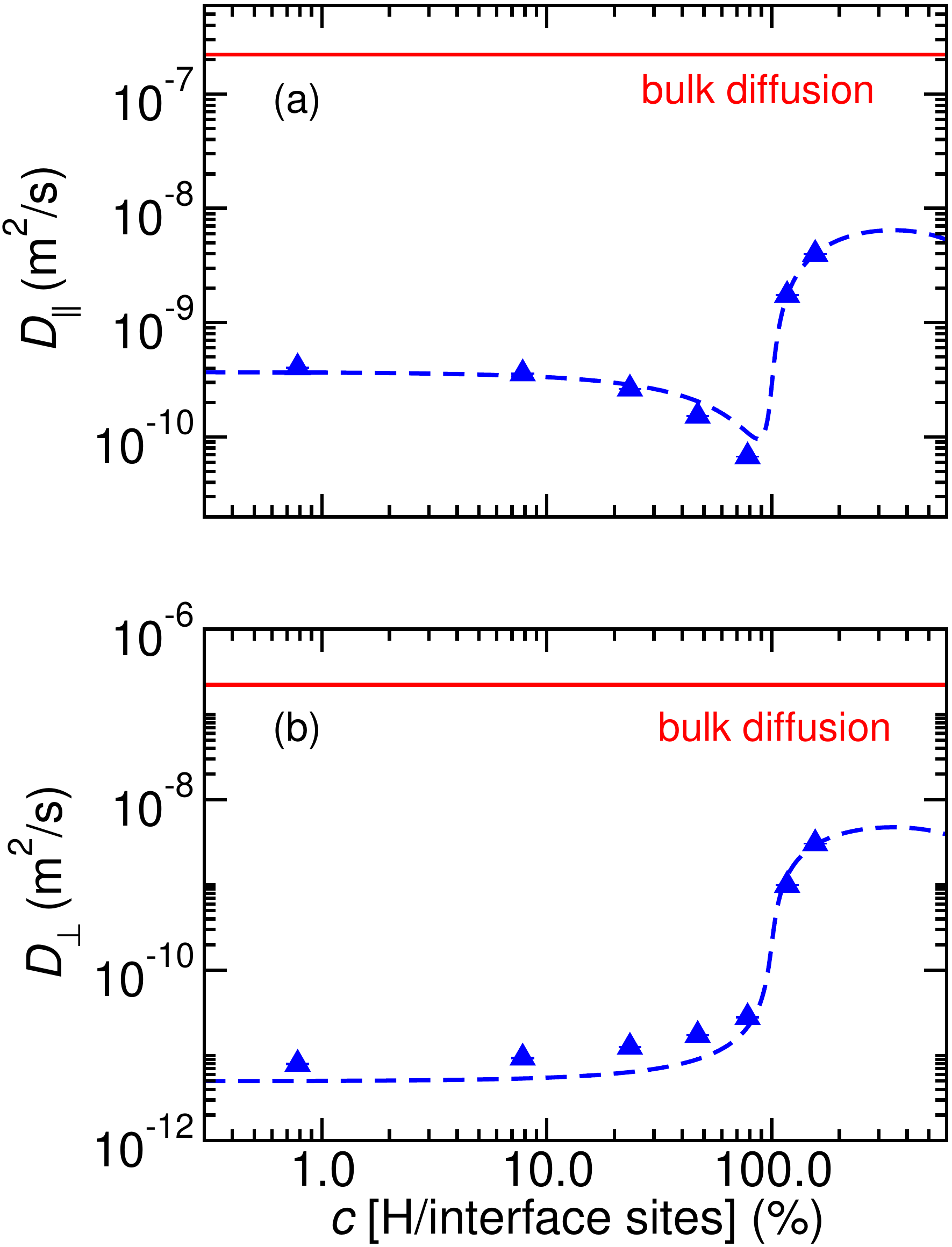}
\caption{(Colour online) Diffusion constants  as a function of  H concentration at $T = 600$~K in an idealised layered structure. The diffusivities parallel  ($D_{\parallel}$) and perpendicular ($D_{\perp}$) to the interface layers are shown in (a) and (b), respectively.  The blue triangles denote the numerical results from the kMC  simulations, the 
dashed lines are the analytic results obtained via Eq.~(\ref{eq:diffgeneral}).  The red curve represents the diffusivity of  H  in perfect bcc-Fe within the dilute limit.
\label{fig:layerHscom}}
\end{figure}

As a second step the dependence of the diffusivity within the layered structure on the H concentration is investigated.
The results of the kMC simulations   are shown in Fig.~\ref{fig:layerHscom} for  $T = 600$~K. The simulation cell  with $z = 22.656 $~\AA~ contains 128  interface sites and 1408 bulk sites.  

Figure~\ref{fig:layerHscom}(a) illustrates the diffusivity parallel to the interface ($D_{\parallel}$).  Similar to the results for the cubic grain structure the diffusivity is constant and lower than bulk diffusion for low H concentrations.  When the number of H atoms approaches the number of available interface sites blocking occurs and a dip is observed for $D_{\parallel}$.  For even higher concentrations the diffusivity increases quickly due to contributions from bulk diffusion.  
The dashed line shows the results of our analytic model obtained from Eq.~(\ref{eq:diffgeneral}).  The fitted geometric prefactors are $\Gamma_{\parallel, {\rm intf}} = 0.455$ and $\Gamma_{\parallel, {\rm bulk}} = 0.638$, and the numeric and analytic results are in very good agreement.  Since the interface layers can be considered as parallel, 2-dimensional planes, the theoretical value for the geometric prefactor is $\Gamma_{\parallel, {\rm intf}} = \frac{2}{4} = 0.5$. Within the bulk region the geometric arrangement is unchanged, i.e.  $\Gamma_{\parallel, {\rm bulk}} = \frac{4}{6} = 0.667$. Again the fitted geometric prefactors are close to the ideal theoretical values.

Figure~\ref{fig:layerHscom}(b) shows the diffusivity perpendicular to the interface plane,  $D_{\perp}$, as a function of H concentration. For low concentrations diffusion perpendicular to the interfaces is much slower than parallel diffusion.  This is due to the fact that the H atoms are largely confined within the interface region.  At large H concentrations bulk diffusion dominates and the diffusivity becomes isotropic.  Since diffusion within the interface layers does not significantly contribute to $D_{\perp}$, there is also no blocking effect observed. 
Instead, the diffusivity  increases smoothly  as the number of H atoms approaches the number of interface sites and bulk diffusion becomes dominant.  
The dashed line is obtained within our analytic model. Also here the agreement between numerical and analytic results is remarkable.  The fitted geometric prefactors are $\Gamma_{\perp, {\rm intf}} = 0.0$ and $\Gamma_{\perp,{\rm bulk}} = 0.470$.  The zero value of $\Gamma_{\perp, {\rm intf}}$ confirms that diffusion within the interface layers does not contribute to $D_{\perp}$.  The value of $\Gamma_{\perp,{\rm bulk}}$ is smaller than the expected value of $0.667$.  The deviation can be explained by the fact that H diffusion in the bulk region perpendicular to the interfaces is interrupted by trapping within the interface planes effectively lowering the overall diffusion constant.  It also indicates that our assumption within our analytic model, that the overall diffusion is a weighted sum over the diffusion within the different regions, does not fully apply to $D_{\perp}$.  Here, the overall diffusion is  a combination of diffusion within the bulk region and trapping within the interface region.  

In a third step we investigate the influence of different types of GBs on the diffusivity.  Our DFT calculations for the $\Sigma 5$ GB in bcc-Fe~\cite{duprbfeh2011} indicate that diffusion perpendicular to the GB interface proceeds via an intermediate site.  Furthermore, the more close-packed $\Sigma 3 [1\bar{1}0](112)$ GB ($\Sigma 3$ in short) in bcc-Fe exhibits rather large diffusion barriers within in the interface region as well as perpendicular to it~\cite{duprbfeh2011}.   We consider three different cases for which the barriers are summarised in Table~\ref{tab:3barrier}.
\begin{table}
\caption{Diffusion barriers, $\Delta E_i$, between various sites $i$ within the idealised layered structure.  The three cases represent three different models to approximate different GBs in bcc-Fe.
 \label{tab:3barrier}}
\begin{ruledtabular}
\begin{tabular}{lccc} 
process $i$   & case (I) & case (II) & case (III) \\ \hline
bulk $\rightarrow$ bulk        			&   $0.088$	& $0.088$	& $0.088$ \\
bulk  $\rightarrow$ intermediate       		&   $0.088$   	& $0.088$	& $0.088$\\
intermediate   $\rightarrow$ bulk       	&   $0.088$  	& $0.288$	& $0.088$\\
intermediate   $\rightarrow$ intermediate      	&   $0.088$    	& $0.200$	& $0.088$\\
intermediate  $\rightarrow$ interface       	&   $0.088$    	& $0.088$	& $0.088$\\
interface  $\rightarrow$ intermediate      	&   $0.600$    	& $0.400$	& $0.500$\\
interface   $\rightarrow$ interface      	&   $0.250$    	& $0.550$	& $0.550$\\
\end{tabular}
\end{ruledtabular}
\end{table}
Case~(I) resembles our original setup where the intermediate and bulk sites are equivalent and the diffusion barrier between interface sites is larger than in the bulk region but still considerably lower than the diffusion barrier out of the interface region.  In case~(II) intermediate sites are introduced where the diffusion barrier between intermediate sites is lower but almost comparable to the escape barrier into the bulk region.
Case~(III) resembles the situation within the $\Sigma 3$ GB in bcc-Fe.  Both the diffusion barrier within and perpendicular to the interface region is rather large.  As already discussed, for a temperature of $T=600$~K and in the dilute limit for case~(I) hydrogen is largely confined to the interface region and almost the entire diffusion takes place in the interface region, although the diffusion barrier is higher than in the bulk region.    
In case~(II) approximately half of all diffusion processes are found in the intermediate region, which only provides about $1/5$ of the available sites.  This indicates that diffusion within the intermediate region is preferred, guiding hydrogen along the grain boundary, thereby partially alleviating the trapping effect of the energetically low lying interface sites.  For case~(III) we find that the interface region only acts as a trap for hydrogen.  Diffusion is almost entirely observed in the bulk region and there is no enhanced diffusivity in the intermediate region, i.e. in this case the GB does not influence the preferred diffusion direction of hydrogen.

Our findings for the idealised layered structures indicate that depending on the nature of the actual GB the interface region might either determine the preferred diffusion direction or trap and effectively immobilise hydrogen atoms at the interfaces.  
In the next Section we extend our layered model to more closely resemble the structure of the $\Sigma 5$ GB in bcc-Fe.


\section{$\maybebm{\Sigma  5}$ Grain Boundary in \lowercase{bcc}-F\lowercase{e}\label{sigma5}}

\begin{figure}
\includegraphics[width=0.2\textwidth]{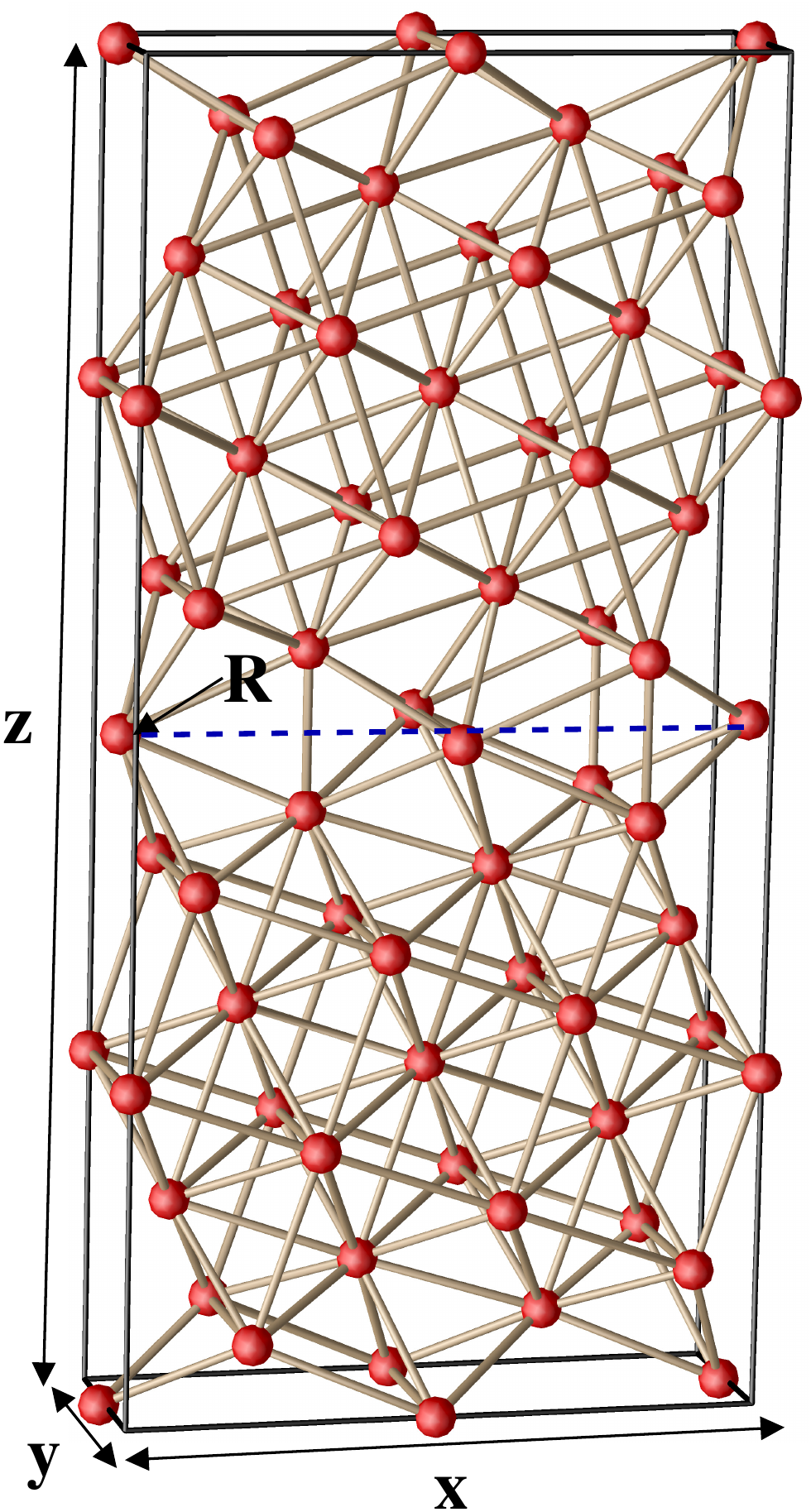}
\caption{(Colour online) Simulation cell of the $\Sigma 5$ grain boundary in bcc-Fe. The red spheres represent Fe atoms, the GB interface is indicated by the blue dashed line.  All interstitial sites are characterised with respect to a reference atom denoted by \sf{R}. 
\label{fig:sig5-40model}}
\end{figure}

The structure of the $\Sigma 5$ grain boundary in bcc-Fe is illustrated in Fig.~\ref{fig:sig5-40model}.  
To setup a kMC model that describes the diffusion of hydrogen within the $\Sigma 5$ GB it is necessary to identify all stable interstitial sites for H atoms within this structure as well as possible diffusion processes between these sites.  Here we performed extensive DFT calculations to obtain reliable values for solution energies and diffusion barriers.
Our results are thus not dependent on any fitted parameters but all input data are extracted from ab-initio calculations.
We have investigated the stability of hydrogen within various interstitial sites in a previous study~\cite{duprbfeh2011} and found that within the $\Sigma 5$ GB interstitial sites close to the interface region are energetically more favourable for H atoms than the tetrahedral site in bulk bcc-Fe.
The DFT calculations revealed   8 distinctive interstitial sites within the $\Sigma 5$ GB.  
The relative solution energies of H within these sites with respect to the most stable site (if3) are summarised in Table~\ref{tab:sitedis}. 

To identify suitable transition states for diffusion processes between the various interstitial sites we employed the nudged elastic band method~\cite{jonsson:1998, jonsson:2000a} as implemented in the VASP code~\cite{vasp1:1993, vasp2:1996}. The computational details can be found in Ref.~\onlinecite{duprbfeh2011}.
We found 12 different diffusion processes, the corresponding barriers extracted from the DFT calculations are summarised in Table~\ref{tab:siterate}.

\begin{table}
\caption{Relative positions of various symmetry-inequivalent interstitial sites within the $\Sigma 5$ grain boundary in bcc-Fe.  The positions are given with respect to the reference atom {\sf R} in Fig.~\ref{fig:sig5-40model}.  Solution energies  are listed relative to the most stable interstitial site, if3. The relative solution energy of a tetrahedral bulk site is 0.500~eV.
\label{tab:sitedis}}
\begin{ruledtabular}
\begin{tabular}{ccccc}
site  &  $\Delta x$ (\AA) &  $\Delta y$  (\AA) & $\Delta z$ (\AA) & $\Delta E$ (eV) \\ \hline
if1  &      7.041 &       0.000   &      0.001 &  0.050 \\
if2  &      3.917 &       0.000   &      0.007 &  0.269  \\
if3  &      5.522 &       0.000   &      0.075 &  0.000 \\
im1  &      4.249 &       0.000   &      0.659 &  0.237 \\
im2  &      1.287 &       0.000   &      0.759 &  0.208  \\
im3  &      7.591 &       0.000   &      1.204 &  0.315 \\
im4  &      6.854 &      -0.001   &      1.504 &  0.373 \\
\end{tabular}
\end{ruledtabular}
\end{table}

\begin{table}
\caption{Hydrogen diffusion barriers for possible transitions between   interstitial sites in the $\Sigma 5$ grain boundary in bcc-Fe.  Values are given  for both the forward and backward process. The process im3 $\leftrightarrow$ im2 takes place between second nearest neighbours, while all other transitions are between  nearest neighbour sites.
\label{tab:siterate}}
\begin{ruledtabular}
\begin{tabular}{lcc}
process      &   $\Delta E_{\rm forward}$  (eV)   &    $\Delta E_{\rm backward}$  (eV)   \\ \hline
if3 $\leftrightarrow$ if1       &   0.118     &      0.068   \\
if1 $\leftrightarrow$ im3       &   0.339     &      0.074    \\
im3 $\leftrightarrow$ im2      &   0.109     &       0.216     \\
im2 $\leftrightarrow$ bulk         &   0.383    &      0.091     \\
if3 $\leftrightarrow$ im4       &   0.429     &    0.056    \\
if2 $\leftrightarrow$ im1       &  0.020     &     0.052    \\
im1 $\leftrightarrow$ if3    &  0.102    &      0.339    \\
im4 $\leftrightarrow$ bulk      &   0.201   &           0.074  \\
bulk $\leftrightarrow$ bulk      &   0.088 &           0.088  \\
if3  $\leftrightarrow$ if3      &   0.250 &           0.250  \\
if3  $\leftrightarrow$ im2      &   0.241 &        0.033     \\
im4  $\leftrightarrow$ im3      & 0.030  &        0.088    \\
\end{tabular}
\end{ruledtabular}
\end{table}

Based on our very detailed DFT study the kMC model of the $\Sigma 5$ GB structure is constructed by mapping the 8 identified interstitial sites onto a lattice connected by the 12 diffusion processes listed in Table~\ref{tab:siterate}.
The $(1 \times 1 \times 1)$ supercell of the $\Sigma 5$ GB shown in Fig.~\ref{fig:sig5-40model}  has the dimensions $x = 8.98$~\AA, $y = 2.84$~\AA, and  $z = 71.78$~\AA.  
In addition to the tetrahedral interstitial sites in the bulk region, the relative positions of the different  interstitial sites close to the interface region with respect to a reference atom {\sf R} (cf. Fig.~\ref{fig:sig5-40model})   are listed in Table~\ref{tab:sitedis}.
The reference atom {\sf R} sits within the GB plane.  Due to the mirror symmetry of the GB interface,  for an interstitial site at ($\Delta x$, $\Delta y$, $\Delta z$)  with $\Delta z > 0.1$~\AA, there is an equivalent site at ($\Delta x$, $\Delta y$, $-\Delta z$).  Since the $\Sigma 5$ GB has a base-centred orthorhombic structure, for an interstitial site  at ($\Delta x$, $\Delta y$, $\Delta z$) there exists an equivalent site at ($\Delta x + x/2$, $\Delta y +  y/2$, $\Delta z$). 
For the kMC simulations we used a model corresponding to a $(2 \times 6 \times 1)$ supercell that is repeated periodically in all three dimensions.  The model contains  48 if1, 48 if2, 48 if3, 96 im1, 96 im2, 96 im3, 96 im4, and 10224 bulk sites.

\begin{figure}
\includegraphics[width=0.5\textwidth]{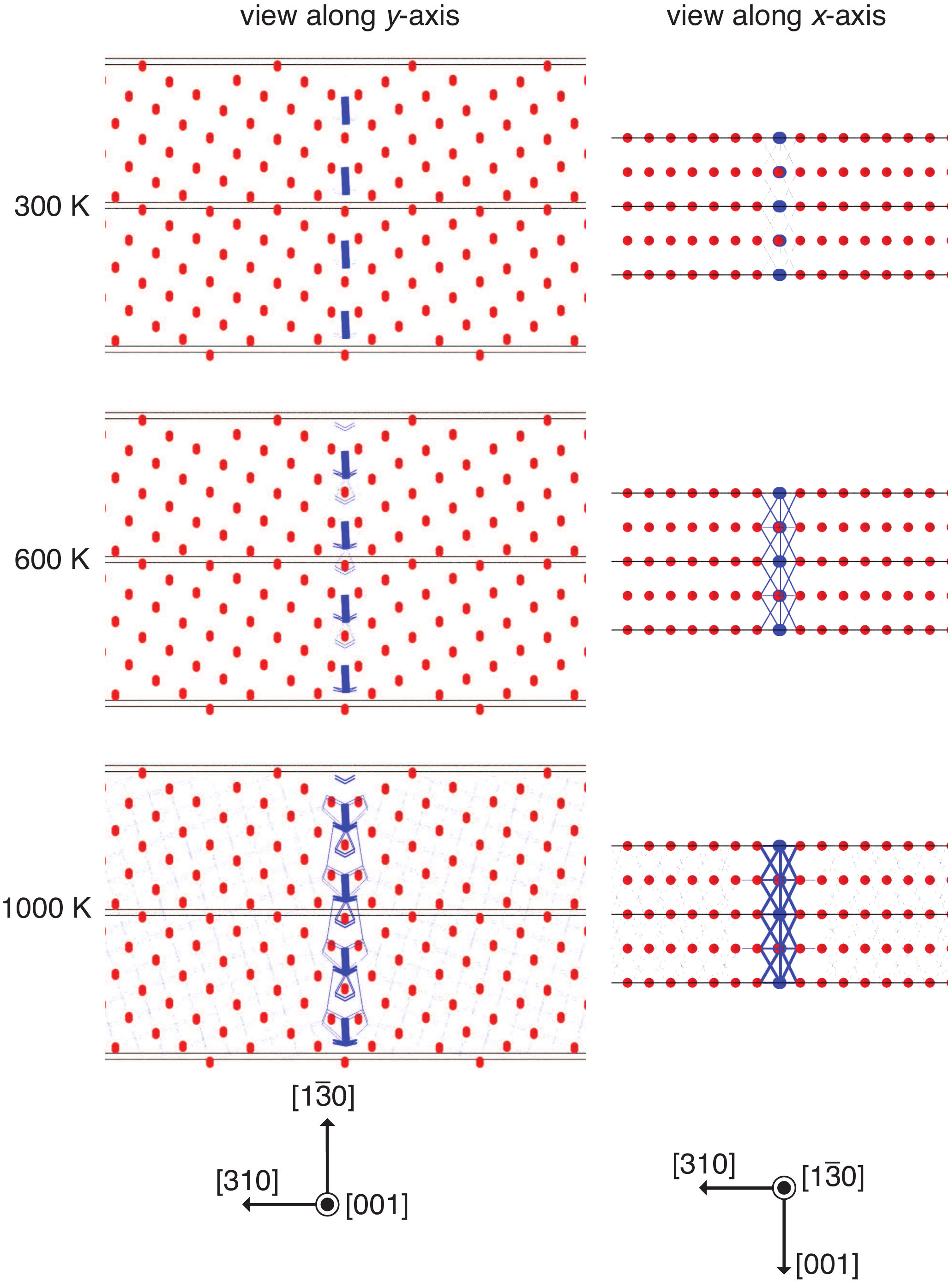}
\caption{(Colour online) Diffusion network within the $\Sigma 5$ GB. The red spheres represent Fe atoms, and the blue lines 
illustrate the diffusion probabilities for various H diffusion paths.
\label{fig:tubes}}
\end{figure}

\begin{figure}
\includegraphics[width=0.4\textwidth]{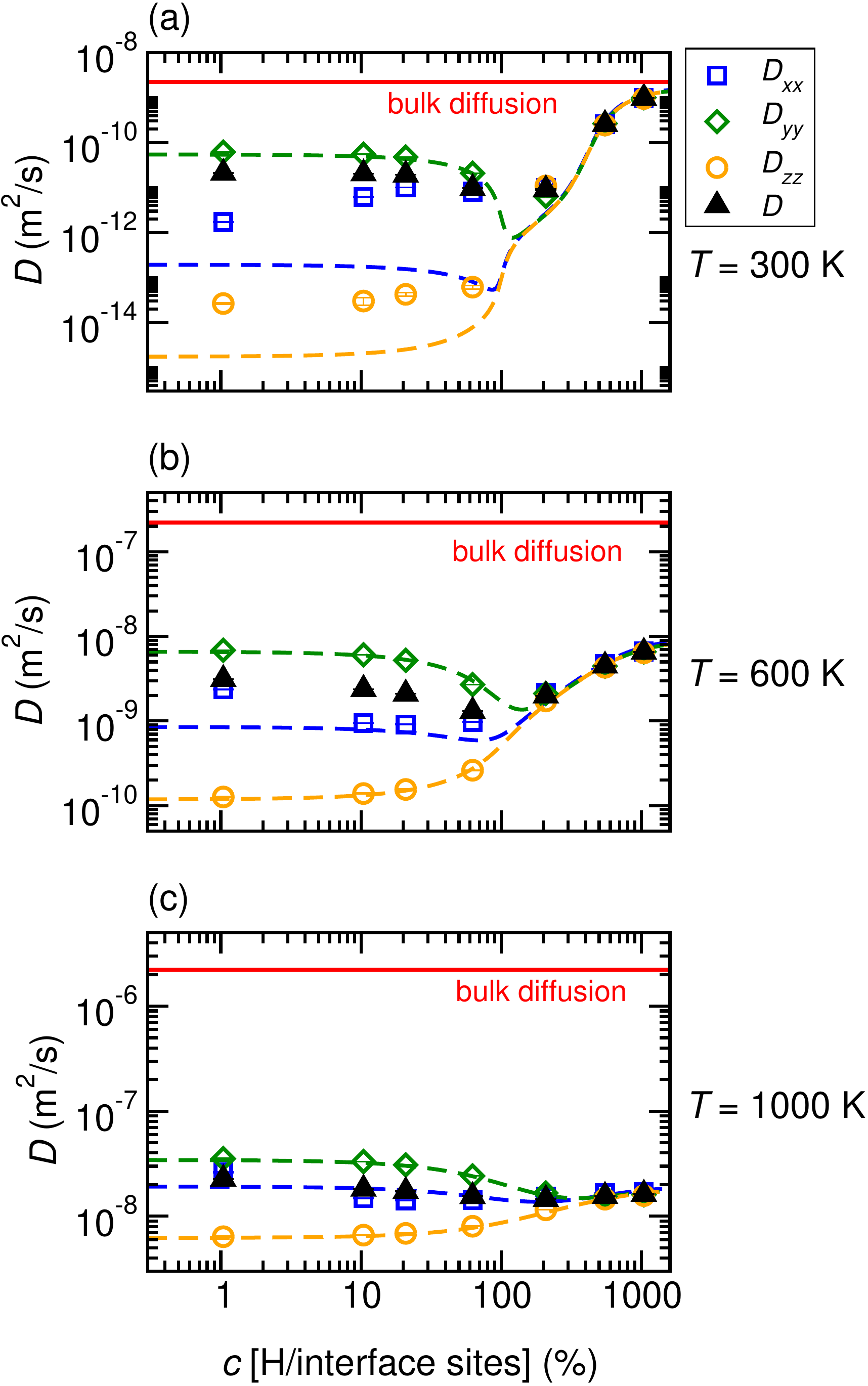}
\caption{(Colour online) Components of the diffusion constant tensor as a function of H concentration  within the $\Sigma 5$ GB in bcc-Fe.  Results are shown for (a) 300~K, (b) 600~K, and (c) 1000~K.  The kMC results for $D_{xx}$, $D_{yy}$, $D_{zz}$, and the overall diffusivity $D$ are represented by blue squares, green diamonds, orange circles, and black triangles respectively.  The  analytic results are shown as dashed lines in the corresponding colour. 
The red curve represents the diffusivity of  H  in perfect bcc-Fe within the dilute limit.
\label{fig:sig5bccdf}}
\end{figure}

The interstitial sites and microscopic diffusion processes create a complex diffusion network for hydrogen that is highly anisotropic.  To illustrate this network the contribution of different processes to the overall diffusivity is shown in Fig.~\ref{fig:tubes} for low H concentrations.  The relative line thickness corresponds to the relative probabilities of observing H atoms diffuse along this connection.  For all three temperatures $T = 300$, 600 and 1000~K diffusion within the bulk region is negligible.  Thus diffusion in $z$-direction is  slow since for this hydrogen has to leave the interface region and cross the bulk region towards the next interface.
The plots on the left side of Fig.~\ref{fig:tubes} show a view along the $y$-axis, $[001]$-direction, illustrating the diffusion network in $x$-direction.  Although there is a thick connection between the if3 and if1 site there exists no continuous diffusion pathway  within the interface.  To diffuse in $x$-direction hydrogen has to leave the interface taking less favourable paths.  Thus diffusivity in $x$-direction is low for small hydrogen concentrations as also shown in Fig.~\ref{fig:sig5bccdf}.  
For the diffusion in $y$-direction hydrogen can continuously move along if3 sites without leaving the interface region as shown in the plots on the right side of Fig.~\ref{fig:tubes} (view along $x$-axis, $[1\bar{3}0]$-direction).
With increasing temperature the diffusion network becomes more isotropic within the interface plane.  This is also reflected in the different components of the diffusion tensor shown in Fig.~\ref{fig:sig5bccdf}.  For $T=300$~K at low H concentrations $D_{xx}$ and  $D_{yy}$ differ by almost two orders of magnitude, whereas at $T=1000$~K they are comparable.

As discussed in the previous sections, for the idealised interface structures the diffusivity shows a strong dependence on the hydrogen concentration.
In Fig.~\ref{fig:sig5bccdf} the diagonal components of the diffusion constant tensor as a function of the hydrogen concentration are shown for $T = 300$, 600, and 1000~K for the $\Sigma 5$ GB.  For low H concentrations diffusion in the interface region dominates and the diffusivity is anisotropic as expected from the discussion of the diffusion network.  For large H concentrations diffusion is again isotropic since  bulk diffusion dominates.
The overall diffusion constants measured in the kMC simulations are the result of the statistical interplay of all possible diffusion processes.  
In our simple analytic model diffusion within a certain site type is associated with one specific barrier, 
whereas  in the $\Sigma 5$ GB there are several processes and thus barriers associated with one site type and  diffusion within the interface layer proceeds via several site types.
It is thus not possible to accurately describe the diffusivity  within our simple analytic model in Eq.~(\ref{eq:diffgeneral}).

Nevertheless, in order to obtain an understanding of how much the diffusion within the GB interface  and the bulk region contribute to the diffusion constant, we approximate the diffusivity within $\Sigma 5$ GB by selected and/or combined processes within the analytic model.  

H diffusion along the $x$-direction, and  within the GB interface follows the path if3$\rightarrow$im4$\rightarrow$im3$\rightarrow$im2$\rightarrow$if3 (cf. Fig.~\ref{fig:tubes}).  We assume that this diffusion path is dominated by the step if3$\rightarrow$im4 with a barrier of 0.429~eV.  
Following Eq.~(\ref{eq:diffgeneral}), we approximate the diffusivity in $x$-direction, $D_{xx}$, as
\begin{eqnarray}
\label{dcxx}
\lefteqn{D_{xx} =} \\ \nonumber   
   &   & p_{\rm if3}(T) D_{0,xx,{\rm if3}}\exp\left(\frac{-\Delta E_{{\rm if3} \to {\rm im4}}}{ k_{{\rm B}}T}\right) \Xi_{\rm if3}(T, N_{\rm H}) \\ \nonumber 
  &   & + p_{\rm bulk}(T) D_{0,xx,{\rm bulk}} \exp\left(\frac{-\Delta E_{\rm bulk}}{ k_{{\rm B}}T}\right) \Xi_{\rm bulk} (T, N_{\rm H})
 \quad ,
\end{eqnarray}
with $D_{0,xx,s} = \Gamma_{xx,s} a_{xx,s}^{2} \nu_{0}$.
The other contributions to the sum are neglected.
Still, as in the case of isolated point defects, the other interstitial sites influence the hydrogen distribution and thus the corresponding probabilities in Eq.~(\ref{dcxx}).  
The hop lengths between two adjacent if3 sites along the $x$-direction is  $a_{xx,{\rm if3}} = 4.486 $ \AA.

Along the $y$-direction, the overall diffusion constant, $D_{yy}$, is similarly approximated within our analytical model.  Diffusion in $y$-direction either takes place within the bulk region or in between if3 sites.
For the hop between two adjacent  if3 sites  along  the $y$-direction, the hop length is $a_{yy,{\rm if3}} = 2.837$ \AA\ and the barrier $\Delta E_{{\rm if3} \to {\rm if3}} = 0.250$~eV, the other diffusion processes are not taken into account.

As discussed in Sec.~\ref{layer}, for H diffusion along the $z$-direction (perpendicular to the interface), the GB interface acts like a point trap.  All geometric prefactors for diffusion between interface sites are essentially zero, and only bulk diffusion contributes to $D_{zz}$.

The analytic results are shown together with the kMC results in Fig.~\ref{fig:sig5bccdf}.
For $D_{yy}$ the agreement is excellent, indicating that diffusion in bulk and along if3 sites are indeed the dominating processes.  For $D_{xx}$ and $D_{zz}$ the analytic results deviate from the numeric data, especially for lower temperatures.  Still, considering the simplicity of our analytic model the agreement is quite remarkable for $T=600$ and 1000~K.

The fitted values of the geometric prefactors are in $x$-direction $\Gamma_{xx,{\rm if3}} = 1.531$, $\Gamma_{xx,{\rm bulk}} = 0.747$, in $y$-direction $\Gamma_{yy,{\rm if3}} = 1.074$, $\Gamma_{yy,{\rm bulk}} = 0.701$, and in $z$-direction $\Gamma_{zz, {\rm bulk}} = 0.705$.  All bulk values are relatively close to the theoretical value of 0.667.  Diffusion in $y$-direction along if3 sites corresponds essentially to a 1D chain with two nearest neighbours, yielding a theoretical value of $\Gamma_{yy,{\rm if3}} = 2/2 = 1$.  The good agreement between fitted and theoretical values for the  geometric prefactors in $y$-direction also indicates that the simple analytic model is suitable here.

Diffusion within $y$-direction, $D_{yy}$, exhibits a small blocking effect as seen within our idealised models, which is again most pronounced for low temperatures and vanishes at higher temperatures.  At $T = 1000$~K there is no dip, but $D_{yy}$ somewhat decreases with increasing H concentration.  This is due to the fact that at 1000~K the effective diffusion constant within the bulk region is actually smaller than in between interface sites, i.e. $D_{yy, {\rm bulk}} < D_{yy, {\rm if3}}$ with $D_{yy} = p_{\rm if3} D_{yy, {\rm if3}} \,\Xi_{\rm if3} + p_{\rm bulk} D_{yy, {\rm bulk}} \,\Xi_{\rm bulk}$, cf. Eq.~(\ref{dcxx}).  The diffusion barrier between if3 sites is larger than between bulk sites, but at  high temperatures the prefactor, 
which is about a factor of 10 larger for diffusion between if3 sites, dominates.  Thus at high concentrations where bulk diffusion has the largest contribution the overall diffusivity, $D_{yy}$, decreases.
At all temperatures, diffusion  perpendicular to the interface planes, $D_{zz}$, is slowest, due to trapping of H within the grain boundary region.  
With respect to diffusion in $x$-direction ($D_{xx}$)  the  kMC results for $T = 300$~K suggest that the if3 site likewise acts as point trap.   At this low temperature H atoms that occupy if3 sites  are nearly immobile in $x$-direction.  Once the H concentration reaches that of if3 sites, diffusion in $x$-direction may occur following the path im2 $\leftrightarrow$ bulk $\rightarrow$ im2 or im4 $\rightarrow$ bulk $\rightarrow$ im4 with smaller diffusion barriers of 0.38~eV and 0.21~eV, respectively.  These effectively lower barriers are also consistent with the increase and relatively large value of  $D_{xx}$ with increasing H concentration as shown in Fig.~\ref{fig:sig5bccdf}~(a). 
For high temperatures and high H concentrations bulk diffusion dominates and the diagonal components of the diffusion tensor are equivalent, indicating an isotropic diffusion behaviour.  In all cases the overall diffusion in the grain boundary structure is slower than diffusion in perfect bcc-Fe bulk.

It is apparent that for the more realistic model of the $\Sigma 5$ GB in bcc-Fe our analytic model is too simple to fully describe the diffusion  of hydrogen.  Nevertheless, the observed trends at high temperatures may be understood qualitatively with a simple analytic approximation.
Analysing the kMC results a diffusion network was established and the processes that dominate the diffusion  under various conditions were extracted yielding a more detailed understanding of hydrogen diffusion within grain boundary structures. 


\section{Conclusions \label{conls}}

Employing kinetic Monte Carlo simulations we have studied hydrogen diffusion within various models representing different arrangements of grain boundaries and point defects in bcc-Fe.  The defect regions exhibit interstitial sites with a significantly lower solution energy for H atoms, effectively acting as trapping sites.  Within an idealised cubic grain structure we observe a characteristic behaviour of the diffusion tensor as a function of hydrogen concentration.  At low concentrations H is confined to the interface region and the diffusivity is low as compared to diffusion in perfect bcc-Fe bulk.  As the number of H atoms approaches the number of interface sites the diffusion constant drops due to blocking of available interstitial sites.  At large H concentrations bulk diffusion dominates the behaviour and a significant increase of the diffusivity is observed.  Additional point defects lower the diffusivity for small H concentrations, but do not change the behaviour for larger concentrations.

Within a layered arrangement of grain boundary planes the diffusion is anisotropic.  Parallel to the interface diffusion  is similar to the one observed within the grain structure.  Perpendicular to the interface diffusion is much slower and can effectively be described by a 1D model of H atoms hopping between neighbouring interface planes.
The effect of the grain boundary on the diffusion of hydrogen strongly depends on the actual solution energies and diffusion barriers between different interstitial sites.

The more detailed model of the $\Sigma 5$ GB in bcc-Fe showed that the overall diffusion  is a complex interplay of various microscopic diffusion processes.  Depending on the conditions (hydrogen concentration, temperature, diffusion direction) different processes dominate the diffusion resulting in a complex diffusion network.  Still, the general trends with respect to temperature and hydrogen concentration may be understood from a simplified analytic model.

We have derived a simple, analytical expression for hydrogen diffusion within microstructures that consist of several distinctive regions (such as bulk, interfaces, point defects).  The analytic model is in very good agreement with the numerical results for the idealised structures.  For the more detailed model of the $\Sigma 5$ GB the analytic model only works for conditions where the diffusion  is  dominated by a few, specific processes, but naturally it fails to capture the more complex interplay between a number of different microscopic diffusion processes.

In all structures the diffusivity is lower than in perfect bcc-Fe bulk, indicating that the grain boundary regions do not serve as fast diffusion channels.  Nevertheless, at low concentrations hydrogen is confined to the interface region, i.e. the arrangement of grain boundary planes and thus the microstructure significantly influences the preferred diffusion direction.

\begin{acknowledgments}
The authors acknowledge financial support through ThyssenKrupp AG, Bayer MaterialScience AG, Salzgitter Mannesmann Forschung GmbH, Robert Bosch GmbH, Benteler Stahl/Rohr GmbH, Bayer Technology Services GmbH and the state of North-Rhine Westphalia as well as the EU in the framework of the ERDF.
\end{acknowledgments}


\end{document}